# Statistics of Night time Ionospheric Scintillation Using GPS data at low latitude ground station Varanasi

S. Priyadarshi[1] & A. K. Singh[2]

[1]Space Research Centre Bartycka 18A Warsaw, Poland

[2]Atmospheric Research Lab., Department of Physics, BHU, Varanasi-221005 India

**Abstract**

Ionospheric scintillation is the rapid change in the phase and/or the amplitude of a radio signal as it passes through small scale plasma density irregularities in the ionosphere. These scintillations not only can reduce the accuracy of GPS/Satellite Based Augmentation System (SBAS) receiver pseudo-range and carrier phase measurement but also can result in a complete loss of lock on a satellite. Scintillation in the ionosphere varies as the sun spot number (SSN), Geomagnetic index (o < Kp < 9), time of year, time of day, geographical position. Most scintillation occurs for a few hours after sunset during the peak years of the solar cycle. Typically delay locked loop/phase locked loop designs of GPS/SBAS receivers enable them to handle moderate amount if scintillations. Consequently, any attempt to determine the effects of scintillations on GPS/SBAS must consider both predictions of scintillation activity in the ionosphere and residual effect of this activity after processing by a receiver. In this work dual frequency ($f1 = 1.5$ GHz, $f2 = 1.2$ GHz) GPS data recorded at low latitude station Varanasi (Geographic latitude $25^0\ 18'$ N, longitude $82^0\ 59'$ E) have been analyzed to monitor the amplitude scintillation index ($S_4$) from January 2009 to December 2009. These analyzed data is used to study the monthly and seasonal variability of ionospheric plasma bubbles at low latitude. We have studied the power spectral density of the signal and using it we have calculated the estimated scintillation pattern velocity.

# 1. Introduction

Ionospheric scintillation on Global Positioning System (GPS) radio links is a phenomenon originating in the Earth's upper atmosphere which has both theoretical and practical interest. Indeed, scintillation is the footprint on radio links of complex plasma dynamics and hence it can be used for remote sensing of ionospheric irregularities Wernik et al. (2003). On the other hand, scintillation may affect radio communications severely, harming their information content and can even cause the loss of the signal by a receiver (Yeh and Liu,1982). In any case, it is desirable to have statistical tools to single out scintillation events and identify their characteristics as effectively as possible. The percentage of irregularities in these fluctuations is usually very small, but it can be as large as nearly 100% near the equator (Basu and Khan, 1976).Variability of ionospheric irregularities are of serious concern to radio communications because these irregularities affect the amplitude and phase of satellite signals. Amplitude variations may induce signal fading, and when depth of fading exceeds the fade margin of a receiving system, message errors are encountered. If navigation is dependent on the Global Positioning System (GPS), then amplitude fluctuations may lead to data loss and cycle slips (Aarons and Basu, 1994). Sudden phase changes may cause a loss of phase lock in GPS receivers (Basu et al., 1995). Equatorial scintillations during a high solar activity period have been found to be sufficiently intense to disable many communication and navigation systems (Groves et al., 1997). Hence, it is necessary to understand the role of space-weather events on scintillations.

Nighttime equatorial F-region is a seat of intense plasma density irregularities encompassing scales from about 1000 km to a fraction of a meter. These plasma density irregularities are associated with the phenomenon of equatorial spread-F as seen in radio soundings of the ionosphere (occurrence of diffuse F-region traces in the ionograms), radio wave scintillation in trans-ionospheric propagation (Banola et al. 2005). It is well established now that large-scale plasma depletions are generated first in post-sunset period through Rayleigh-Taylor Instability and the depleted regions rise fast to cover the entire F region including topside ionosphere. Plasma processes then give rise to instabilities acting on the steep gradients available and generate smaller and smaller scale sizes in a cascade process (Haerendel, 1973). Since the equatorial spread-F irregularities extend along the field lines, higher the altitude of plasma depletion, wider the latitudinal extent of the irregularities. Although the ionospheric delay is the biggest error source for GPS navigation, it can be directly measured by future dual frequency GPS avionics and higher order ionospheric errors are not problematic for Localizer Performance with Vertical guidance i.e. LPV-200 (Dutta-Barua et al. 2008). However deep and frequent GPS signals fade due to electron density irregularities in the ionosphere raise a concern about the operational availability of LPV-200. Trans-ionospheric radio waves interfere constructively and destructively when they pass through electron density irregularities. This phenomenon can be understood as multipath inside the ionosphere. As a result a receiver experiences amplitude fading and phase jitter of the received signal. This phenomenon referred to as ionospheric scintillation (Crane 1977, Gwal et al., 2004). The signal fades are rapid with decorrelation times of several seconds or less depending on scintillation intensities (Basu et al., 1985).Many workers (Aarons et al., 1980; Rastogi et al., 1981, 1990; Das Gupta et al., 1985; Pathan et al., 1991; Kumar et al., 1993) have been studied the geomagnetic control over the occurrence of scintillations at equatorial and low latitude. Aarons et al., 1980 and Rastogi et al., 1981observed that with the increase in magnetic activity the probability of scintillation increases during the post midnight period. These observations are true in all the latitude sectors, but season as well as longitude affects the pre-midnight phenomenon. Rastogi et al., 1990 have studied the effects of geomagnetic disturbances on scintillations in India and American zone.

In this work we have studied the occurrence of night time amplitude scintillation for year 2009 using GPS data of ground station BHU Varanasi (Geographic latitude $25^0 \, 18^/$ N, longitude $82^0 \, 59^/$ E). This study shows the characteristic of equatorial low latitude ionospheric scintillation in summer, winter and equinox. To quantify the scintillation and to show the information about the spectral content of irregularities causing scintillation, we have plotted power spectrum of amplitude scintillation index ($S_4$) for selected events in year 2009. The roll-off and 'Fresnel' frequencies have been identified for each of the individual spectrum computed. This is felt a better method of comparison of spectral slopes and de-correlation distances. Autocorrelation analysis is another way of characterizing the scintillation fading. It is also an effective means of achieving time diversity improvements. Autocorrelation interval is defined as the time lag for which the level of correlation of the signal decreases to 50% of its maximum value (Umeki *et al.*, 1977). The correlation interval multiplied by the velocity transverse to the ray path would provide the de-correlation distance on the ground. We have plotted autocorrelation function for the same specific events in year 2009.

## 2. Data and Method of analysis

Employing a GPS receiver we have monitored scintillation (S4) index at Varanasi (Geo. lat. $25.3^0$ N, long. $82.99^0$ E), India since January 2009. In this work we have taken the condition if S4 index is greater than 0.2000 then there is ionospheric scintillation. We have studied the occurrence of ionospheric scintillation events for one complete year, starting from Jan. 2009 to December 2009.

We used GSV400B GPS receiver. The purpose of the GSV400B GISTEM is to collect ionospheric scintillation and TEC data for all visible GPS satellite(up to 10), and up to 3 SBAS-GEO satellites, and output data logs, called ISMRB or ISMRA, to a serial port in either ASCII or binary format. Either of two (NovAtel GPSolution4 or SLOG) programs can be can be used to control the GSV400B operations, but SLOG is recommended for collecting scintillation logs.

Ionospheric scintillation measurement was performed using the GPS Ionospheric Scintillation/TEC Monitor (GISTM), model GSV4004 (Van Dierendonck et al. 1993). The system is NovAtel's Euro4 dual-frequency receiver version of the OEM4 card with special firmware, which was developed to maintain lock even under strong scintillation conditions. The amplitude scintillation was monitored by computing the S4 index, which is defined as the standard deviation of the received signal power normalized to the average signal power. It is calculated for each 1-minute period based on a 50-Hz sampling rate. The GISTM also computes the S4 index due to ambient noise in such a way that a corrected S4 index (without noise effects) can be computed (Van Dierendonck et al. 1993). A series of experiments were set up to study the effects of cutoff frequency for amplitude and phase filtering (Forte 2005). Comprehensive discussion about the effects of filtering parameters on scintillation can be found in Forte and Radicella (2002, 2004).

The Corrected S4 is obtained by differencing the S4 Correction from the Total S4 is an RSS sense. If the S4 Correction is larger than the Total S4, simply set the corrected S4 to 0, since the S4 value is obviously due to noise.

## 3. Results:

### 3.1 Statistics of Scintillation Study:

Here in the present work we have taken the numerical value of S4 index 0.2000 as the threshold value of the ionospheric amplitude scintillation. In this paper we have studied only the night time scintillation. So we have analyzed data from 17:00:00 LT to 06:00:00 LT hours.

We have started our study with the study of the amplitude scintillation in summer 2009. Here we observed that the occurrence of amplitude scintillation peaks at 18:30:00 local time hours in pre mid night hours, we want to report that we observed more than one peak in pre-midnight time while only one peak at 03:30:00 local time hours in post mid night period as shown in FIGURE 1. Occurrence of the amplitude scintillation during winter 2009 observation shows the peak at 21:00 LT (Pre-midnight) and at 24:00 LT (exactly at and after midnight). After the post midnight the number of occurrence decreased rapidly and it starts rising slowly at 05:30 LT. It can be seen in FIGURE 2 while in the same study for the equinox 2009 has been carried out. We observed the pre-midnight peak at 22:00:00 local time hours while two peaks at 01:00 LT and 05:00 LT after post midnight as in FIGURE 3. FIGURE 4 represents the contour plot for the occurrence of amplitude scintillation for year 2009. Here we have observed the maximum occurrence in Jun-July and August during 17:00 – 19:00 local time hours. In month of December we observed minimum number of occurrence overall.

In such statistical study it is very important to see the percentage occurrence of amplitude scintillation in different months. FIGURE 5 is the bar plot for the percentage occurrence of amplitude scintillation for the year 2009. This shows the percentage occurrence is maximum for the month of February (i.e. 12.65 %) and minimum for December (4.16 %). Then we have characterized the amplitude scintillation according to the module of their amplitude scintillation index. Here we have divided the number of occurrence in three categories: (**I**) S4~ 0.20 – 0.25, (**II**) S4~0.25 – 0.30, (**III**) S4 > 0.30. In this plot it is clear that maximum number of scintillation of category (**I**) is observed in January, February and March 2009. Their number of occurrence is 113, 140 and 125 respectively while minimum in month of December 2009 (49 scintillation events). Category (**II**) scintillation occurrence is maximum during May and June 2009 number of scintillation events is 14 and 15 respectively while minimum in December.

Category (**III**) scintillation i.e. scintillation whose S4 index is greater than 0.30 are observed maximum in January and minimum from June to September 2009 as shown in FIGURE 6.

The irregularities at low latitudes are observed at irregular intervals and their duration increases as one move closer to the geomagnetic equator. The irregularities observed during summer months are relatively weak as compared to those recorded during winter and equinox months. The electric field at the site plays a dominant role in shaping the development of these irregularities. This is well known that the field is eastward during the day and is westward after sunset. Before this direction reversal, there is a sudden enhancement of the eastward electric field. The field and its variation is season and solar activity dependent. Any change in the electric field influences the occurrence of low-latitude scintillations.

### 3.2 Scintillation power spectra computed from the raw data:

Amplitude scintillation power spectra computed from the raw data for year 2009. The raw data were detrended off-line with the sixth-order Butterworth filter having the cut-off frequency at 0.05 Hz. We have adopted Fremouw *et al.* (1978), method to determine the trend. Figure 7 shows the example of the power spectra computed for the raw data to study the irregularity velocity for amplitude scintillation. From the raw data, the amplitude scintillation power spectra can be computed. An example of such a calculation for the specific cases of year 2009 is shown in figure 7. The spectral indices *p* is computed by the least squares fit. One should note, however, that the frequency interval over which the fit makes sense is much shorter for the amplitude. A maximum around 1 Hz on the amplitude spectrum is identified with the Fresnel maximum. on the amplitude spectrum is identified with the Fresnel maximum. The Fresnel frequency *fF* can be computed by the least squares fit of a parabola and used to estimate the scintillation pattern velocity

$$\mathbf{V} = (\lambda\, z)^{0.5} \cdot fF$$

provided a slant distance *z* to the irregularity layer is assumed or known from other sources. Assuming the irregularity layer height of 350 km, for the elevation angle of 45 degrees, $\lambda = 0.19$ m, and *fF* = 1 Hz, the pattern velocity estimated from the spectrum shown in Fig. 7 is about 200 m/s.

## 4. Discussion

Ionospheric scintillations were initially observed on the amplitude recording of signals from discrete radio sources in the sky and the phenomena were suggested to be associated with spread-F (Wild and Roberts 1956, Koster 1958, Bhargava 1964). The F-region irregularities generally associated with the spread-F have been considered to be the main cause of intense night-time scintillations (Huang 1970, Chandra and Rastogi 1974, Chandra et al. 1979, Rastogi 1982; Dabas and Reddy 1986; MacDougall 1990). Scitillations of trans-ionospheric signals for very high frequency (VHF) range have been studied at equatorial low latitudes (Rastogi et al. 1977; Huang 1978; Krishnamoorthy et al. 1979; Aarons 1982; Rastogi et al. 1982; Rastogi 1983; Vijay Kumar et al. 1988; Pathan et al. 1991; Rastogi et al. 1991). Most of the major causes of satellite communication outages are satellite signal fading due to many factors such as climatically, atmosphere and amplitude scintillation. That cause of amplitude scintillation fluctuation is enhancement and fading rapidly, it has result in worsening of accurate data. The study of the amplitude scintillation occurrence, the diurnal and seasonal variation is importance information for obtaining the communication planning via satellite link or system margin design appropriately to the real condition under scintillation effect in each station. At low latitude at the time of sunset an enhanced eastward electric field called the pre-reversal enhancement generally developed at the F-regions heights. As a result, the ionosphere moves upward, develops steep density gradients in the bottom side F-region and becomes unstable to the Rayleigh-Taylor instability (Kelley, 1989). Large scale plasma depletions, called plasma bubbles, form in the bottomside of the F-region and becomes populated with small-scale irregularities as the bubbles rise to great heights. The small-scale irregularities causes intense scintillations of satellite signals along two belts of high ionization density at 20 degree N and 20 degree S geographic latitudes, called the equatorial anomaly (Basu et al., 1988). Scintillations at low latitudes, as in other regions, attain a maximum value during solar maximum period owing to the increased value of the background ionization density. However, low latitude scintillation is not fully dictated by solar transients, such as magnetic storms. In this region, intense scintillations are observed during magnetically quite period as well. When a satellite based communication or navigation system operates in a scintillating environment and loss of signal occurs, an average user attributes the problem to interference or to failure of equipment either at the transient end, the receiving end or satellite itself. If forecasts of scintillations can be provided, the resources of the users will not be misused and instead may evolve alternate strategies. Even though dealing with these strategies is quite interesting but this is beyond the scope of the present work. We have shown here the statistical variation in the occurrence of night time ionospheric scintillation in different seasons for the year 2009.

**6. Conclusion**

Analyzing the GPS data of the year 2009 for studying amplitude scintillation we have following points of remark:

1. In summer, winter and equinox the maximum numbers of amplitude scintillation events are observed in February and minimum in December 2009.
2. In summer, winter and equinox we got peak of the amplitude scintillation events at **18:30:00, 21:00:00** and **22:00:00** in pre-midnight local time hours respectively, while at in post mid-night period it is **03:30:00**, **00:00:00** and **01:00:00** local time hours respectively.
3. We have observed the maximum number of scintillation having amplitude scintillation index ranging from 0.20 to 0.25 and minimum scintillations whose S4 index is greater than 0.30. From this observation we are concluding here that at our location we have frequent occurrence of weak ionospheric scintillatons while very few intense night time low latitude scintillations.
4. In summer we observe the peak of occurrence just after the sunset but in winter and equinox the premidnight peak sifts to the summer peak time. More shifts is observed in equinox.
5. We have observed the equatorial Plasma bubbles of duration 70 to 143 minutes.

**Caption to the figure:**

Figure: 1 Occurrence of amplitude scintillation in summer 2009.

Figure 2: Occurrence of amplitude scintillation in winter 2009.

Figure 3: Occurrence of amplitude scintillation in equinox 2009.

Figure 4: Contour Plot for the annual study of the amplitude scintillation for the year 2009

Figure 5: Percentage occurrence of amplitude scintillation in year 2009

Figure 6: Magnitude wise occurrence of amplitude scintillation in year 2009

Figure 7: Amplitude scintillation power spectra computed from the raw data. Dashed, straight lines show the least squares fits over the frequency interval indicated by the end-points of lines. For the clarity of presentation lines are shifted vertically. The spectral index is approximately 2.

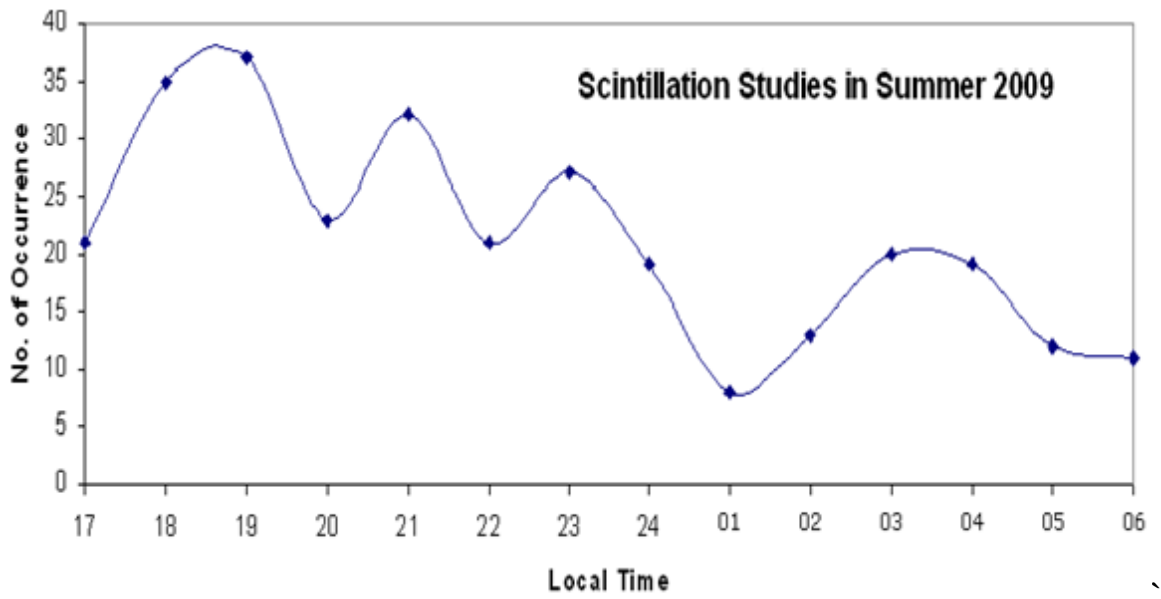

Figure 1

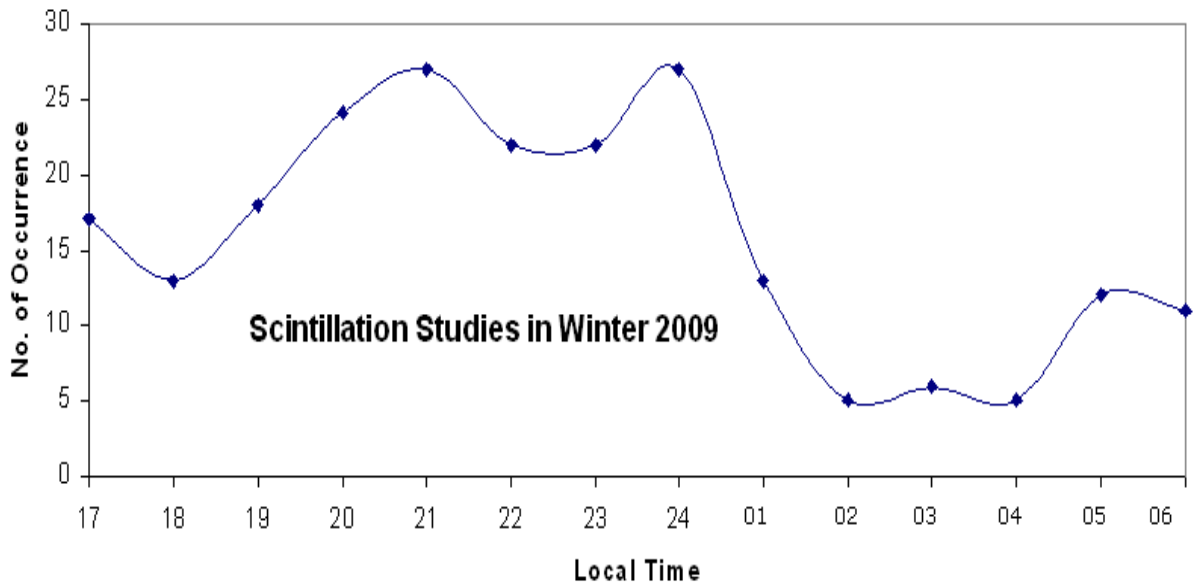

Figure 2

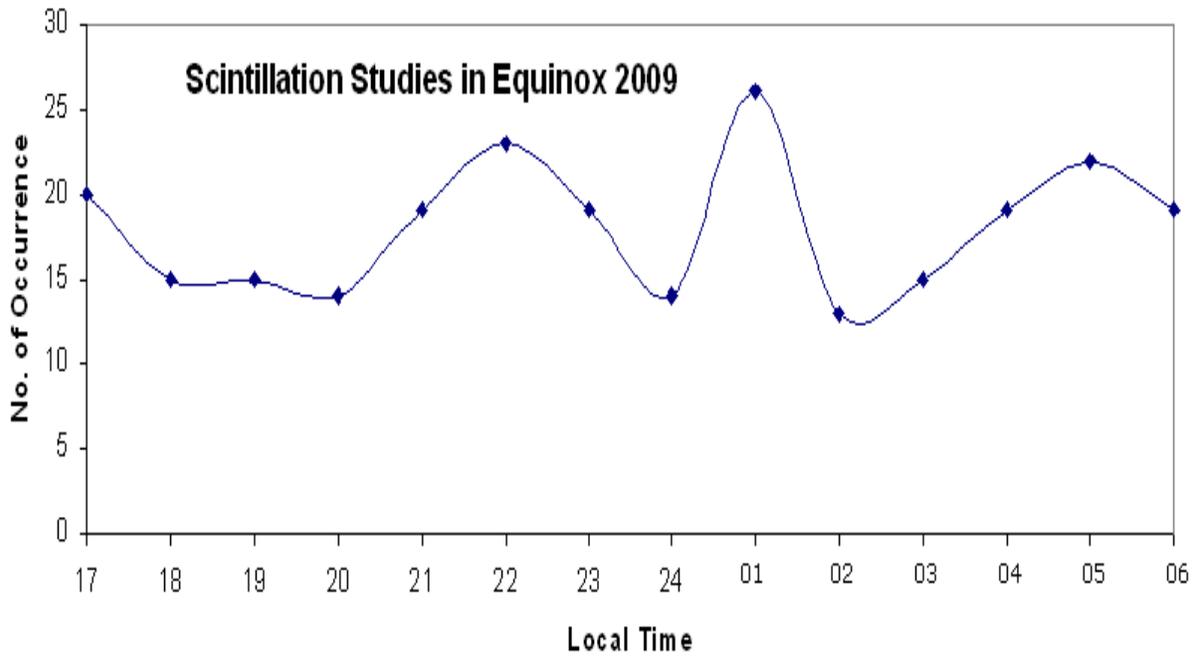

Figure 3

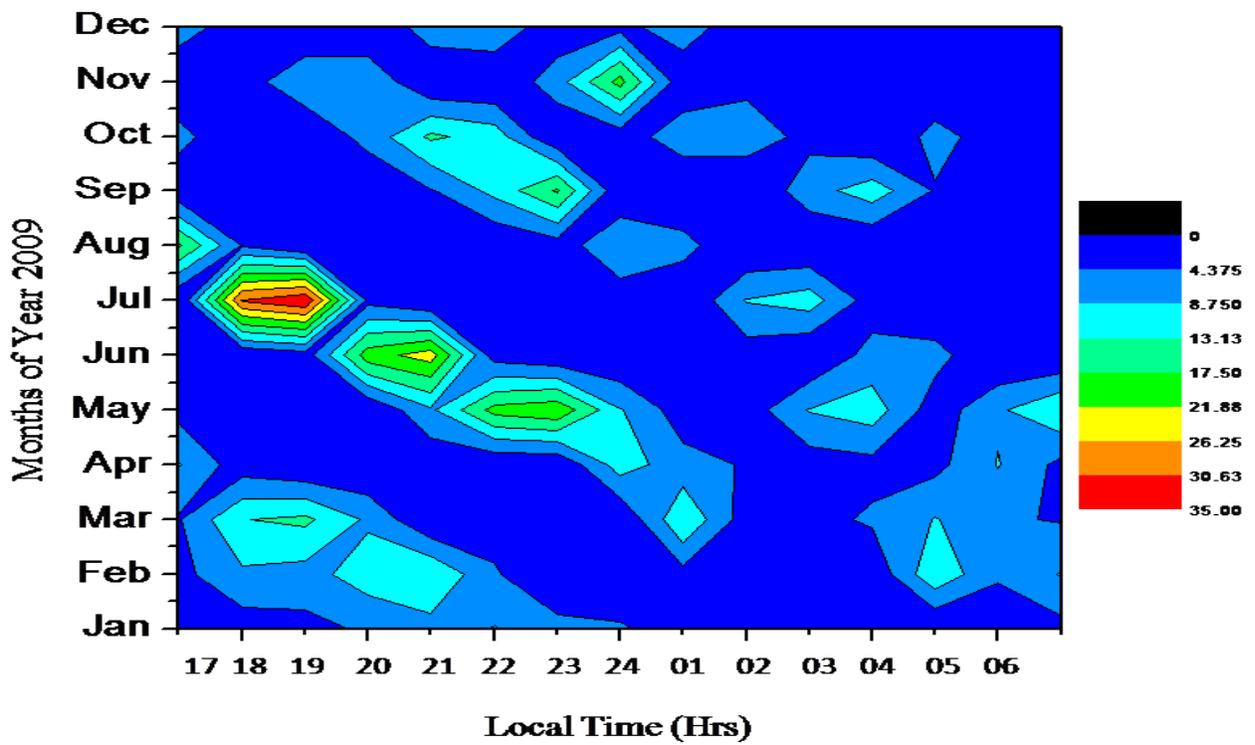

Figure 4

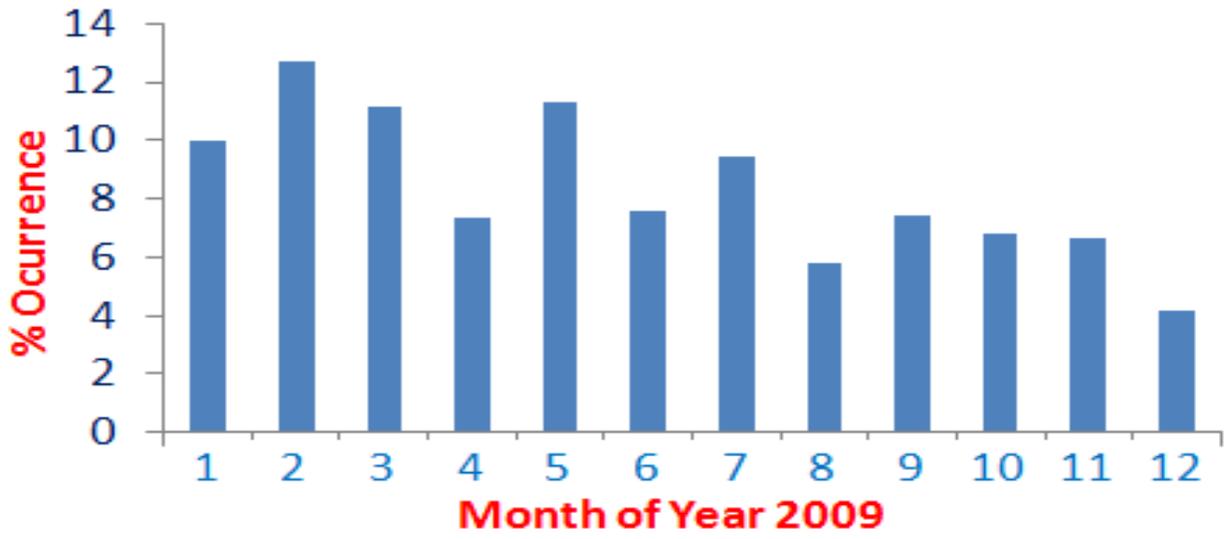

Figure 5

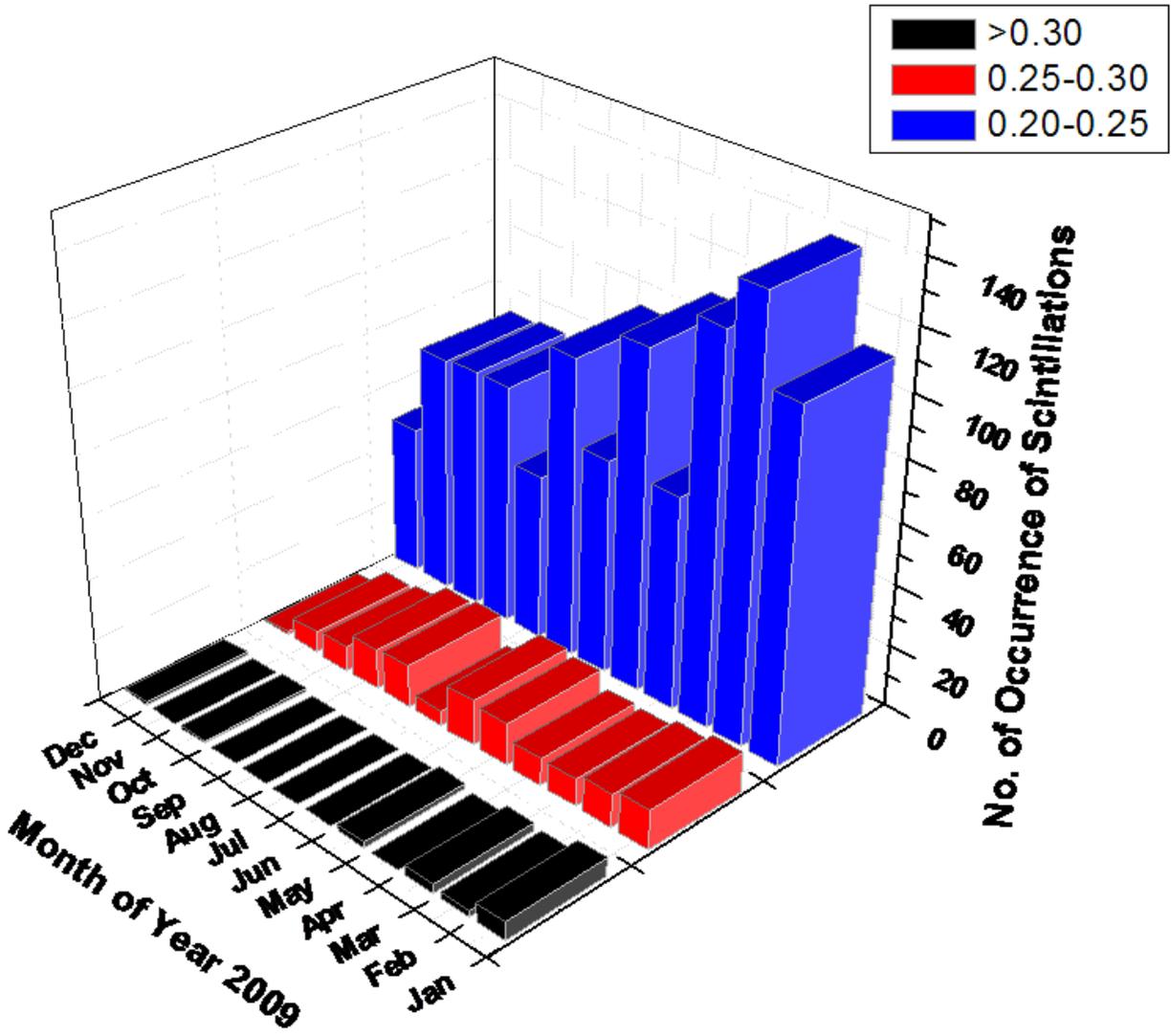

Figure 6

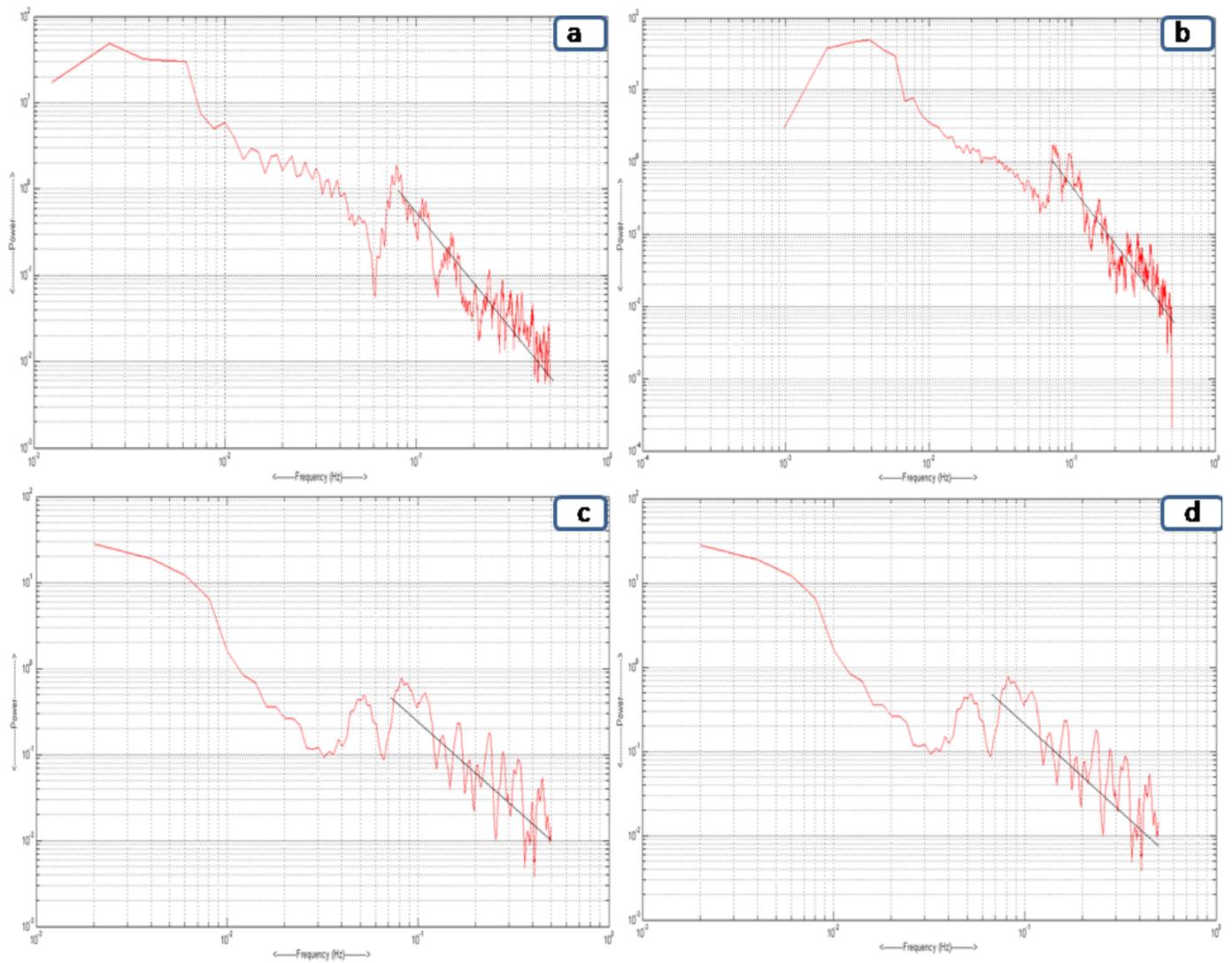

Figure 7.